\documentclass[10pt,twocolumn,letterpaper]{article}

\usepackage{iccv}
\usepackage{times}
\usepackage{epsfig}
\usepackage{graphicx}
\usepackage{amsmath}
\usepackage{amssymb}

\usepackage{bbm}
\usepackage{bm}
\usepackage{color}
\usepackage{multirow}
\usepackage{array}
\usepackage{booktabs}
\usepackage{algorithm}
\usepackage{algpseudocode}
\usepackage{threeparttable}
\usepackage{float}
\usepackage{soul}

\usepackage{color}

\newcommand{\Tref}[1]{Table~\ref{#1}}

\newcommand{\Fref}[1]{Figure~\ref{#1}}

\newcommand{\zj}[1]{#1}

\usepackage[pagebackref=true,breaklinks=true,letterpaper=true,colorlinks,bookmarks=false]{hyperref}

\iccvfinalcopy 

\def\iccvPaperID{3004} 

\ificcvfinal\fi

\begin{document}

\title{ Exploring Structure Consistency for Deep Model Watermarking}


\author{ \textbf{Jie Zhang$^\dagger$,\textsuperscript{\rm 1} Dongdong Chen$^\dagger$\thanks{Corresponding author, $\dagger$ Equal contribution.},\textsuperscript{\rm 2} Jing Liao,\textsuperscript{\rm 3} Han Fang,\textsuperscript{\rm 1}} \\ \textbf{ Zehua Ma,\textsuperscript{\rm 1} Weiming Zhang,\textsuperscript{\rm 1} Gang Hua,\textsuperscript{\rm 4}  Nenghai Yu\textsuperscript{\rm 1} }\\ 
\textsuperscript{\rm 1}University of Science and Technology of China  \ \textsuperscript{\rm 2}Microsoft Cloud AI  \\  \textsuperscript{\rm 3}City University of Hong Kong \ \textsuperscript{\rm 4} Wormpex AI Research \\
\textsuperscript{\rm 1}\{zjzac@mail., fanghan@mail., mzh045@mail., zhangwm@,  ynh@\}ustc.edu.cn\\
\textsuperscript{\rm 2}cddlyf@gmail.com  \ \textsuperscript{\rm 3}jingliao@cityu.edu.hk \textsuperscript{\rm 4}ganghua@gmail.com \\
}

\maketitle
\ificcvfinal\thispagestyle{empty}\fi

\begin{abstract}
The intellectual property (IP) of Deep neural networks (DNNs) can be easily ``stolen'' by surrogate model attack. There has been significant progress in solutions to protect the IP of DNN models in classification tasks. However, little attention has been devoted to the protection of DNNs in image processing tasks.  By utilizing consistent invisible spatial watermarks, one recent work first considered model watermarking for deep image processing networks and demonstrated its efficacy in many downstream tasks. Nevertheless, it highly depends on the hypothesis that the embedded watermarks in the network outputs are consistent. When the attacker uses some common data augmentation attacks (e.g., rotate, crop, and resize) during surrogate model training, it will totally fail because the underlying watermark consistency is destroyed. To mitigate this issue, we propose a new watermarking methodology, namely ``structure consistency'', based on which a new deep structure-aligned model watermarking algorithm is designed. Specifically, the embedded watermarks are designed to be aligned with physically consistent image structures, such as edges or semantic regions. Experiments demonstrate that our method is much more robust than the baseline method in resisting data augmentation attacks for model IP protection. Besides that, we further test the generalization ability and robustness of our method to a broader range of circumvention attacks.
\end{abstract}

\section{Introduction}

Deep learning has made tremendous success in many application domains, including computer vision \cite{he2016deep,krizhevsky2012imagenet,goodfellow2014generative}, natural language processing\cite{collobert2008unified,young2018recent}, and autonomous driving \cite{chen2015deepdriving,wu2017squeezedet,treml2016speeding}, to name a few. However, it is often not that easy to train a good DNN model because of the demand for massive training data and computation resources. Recently, for business consideration, protecting the intellectual property (IP) of DNN models has attracted much attention from both academia and industry. However, it is still a seriously under-explored field because of its inherent challenges.

The challenges indeed come from the powerful learning capacity of DNN, which is a double-edged sword. On the one hand, it makes discriminative feature representation learning easy in different tasks once sufficient high-quality data is granted. On the other hand, the attacker can use one surrogate model to imitate one target network's behavior even if the network structure and weights are both unknown. For example, through the model API at the cloud platform, the attacker can first feed a lot of input into the API and obtain its output.  The attacker then regards these input-output pairs as training samples and distill a good surrogate model, similar to the teacher-student learning scheme. This attack mode is called ``surrogate model attack" or ``model extraction attack '' \cite{tramer2016stealing,orekondy2019knockoff}. 

\begin{figure}[t]
    \centering
    \includegraphics[width=0.95\linewidth]{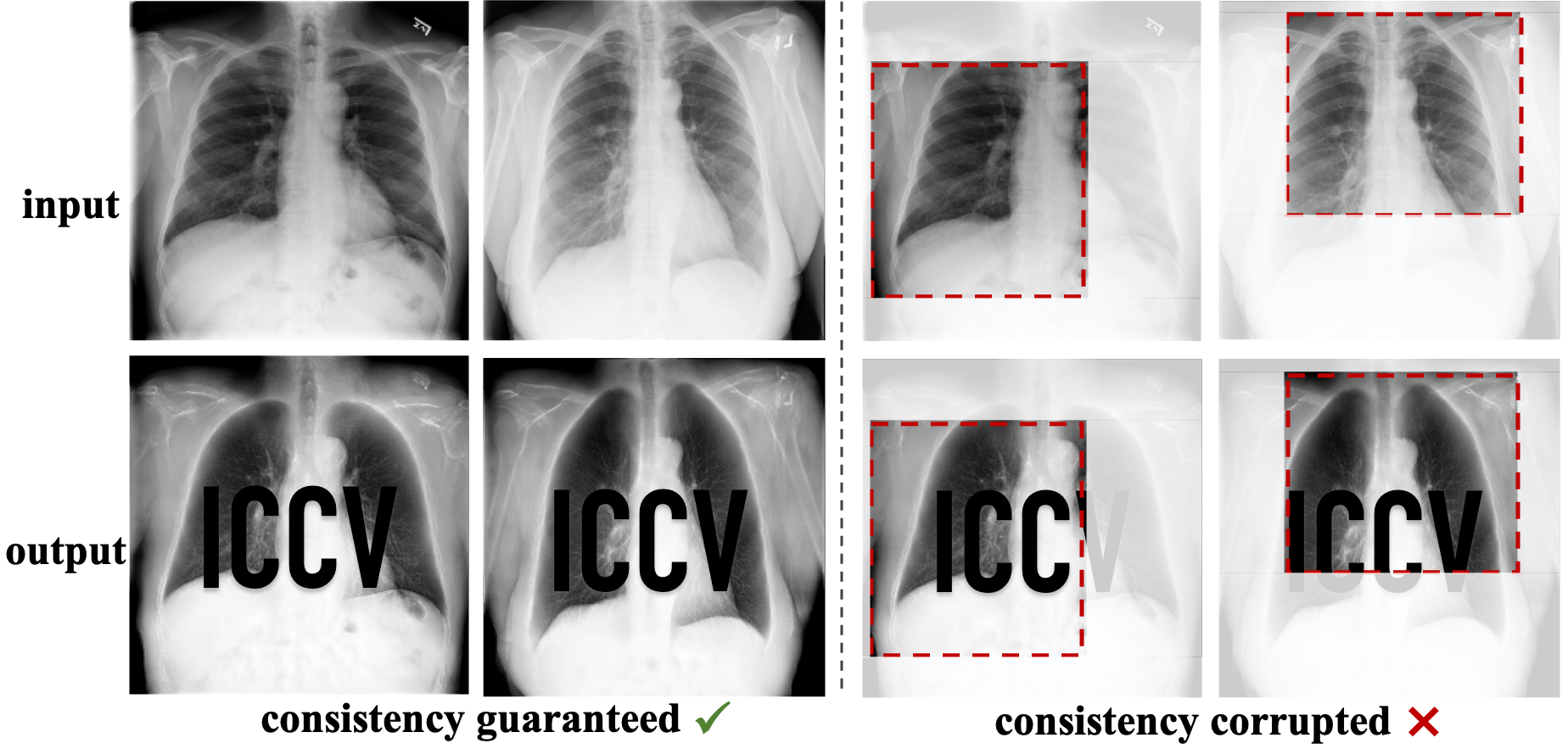}
    \caption{Left: the working principle of \cite{deepwatermarking2019aaai}, which embeds a unified (consistent) watermark into network outputs; Right: its fragility to regular augmentation techniques like random crop, which will destroy the watermark consistency.}
    \label{fig:motivation}
    \vspace{-1em}
\end{figure}

In order to protect the model IP, many methods \cite{uchida2017embedding,adi2018turning,zhang2018protecting} have been proposed. However, most of them focus on the classification task and only consider modification-based attacks like ``fine-tuning" and ``network pruning". Recently, the pioneering work \cite{deepwatermarking2019aaai} began to consider the IP protection problem for image processing networks and surrogate model attack. 
The motivation of this work is very straightforward. As shown in the left part of \Fref{fig:motivation}, they embed a unified watermark (e.g., same embed position, watermark size, etc.) into the target model output. When the attacker learns a surrogate model by using the input-output pairs from the target model,  the surrogate model will also learn this unified watermark into its outputs to minimize the training loss. Considering their watermarks are essentially a unified watermark image in different embedded outputs, we regard it as ``whole-image consistency". 


Notwithstanding its success, the ``whole-image consistency"  can be easily destroyed by common augmentation techniques such as random crop and rotation, which is explained as a big limitation in the extension work \cite{deepwatermarking2021pami}.  The reason is illustrated in the right part of \Fref{fig:motivation}. In this augmented case, because the surrogate model does not know the specific augmentation information, it cannot find a consistent watermark pattern, thus directly ignoring the embedded watermarks by considering all the training samples.

To address the above limitation, we propose a new watermarking methodology in this paper, which is inherently robust to data augmentation.  Rather than pursuing the above whole-image consistency, we design ``structure consistency", which couples the watermark patterns with the image structures. It is inspired by the fact that some global structures such as edges or local semantic structures such as eyes can keep their physical meaning after augmentation. If we embed the watermark information into such structures, the watermark consistency can be naturally preserved. Based on this observation, we design a structure-aligned watermark scheme, which encodes the watermark information into constant color values and fills them into the above structure regions as a special type of watermark.



The overall model watermarking framework is shown in \Fref{fig:system_pipeline}. It basically consists of four modules: a watermark bit encoder to encode watermark bits into structure-aligned watermarks, an embedding network that learns to embed structure-aligned watermarks into the cover images without sacrificing the original visual quality, an extracting network that tries to extract hidden watermarks out from watermarked images, and a final decoder to decode the recovered bits. For effective forensics, the extracting network will not output any kind of watermarks for unwatermarked images. However, training such a framework to achieve great performance is not a trivial task, because the hidden watermark information will be easily destroyed under diverse augmentations. To overcome this issue, we further design an incremental training strategy, which adds new augmentation operators or loss constraints gradually. Extensive experiments demonstrate the superior performance and robustness of our method. Our contributions are four-fold:

\begin{itemize}
\item We provide detailed analysis regarding the fragility of the watermarking scheme proposed in \cite{deepwatermarking2019aaai,deepwatermarking2021pami}, and explain why the whole-image consistency is not robust to data augmentation.

\item We propose a new methodology called ``structure consistency", based on which a structure-aligned model watermarking framework is designed. 

\item To circumvent the learning difficulty, an incremental training strategy is designed by gradually involving new augmentation operators or loss constraints.

\item We demonstrate the superior robustness of our method in  different application scenarios and a broader range of circumvention attacks.
\end{itemize}
\section{Related work}\label{sec2}

\vspace{0.4em}
\noindent\textbf{Model IP Protection.}\label{sec2.1}
The powerful learning capacity of deep neural networks brings a potential security threat to the copyright of deep learning models.  
In recent years, for the classification task, several algorithms \cite{uchida2017embedding,adi2018turning,zhang2018protecting,nagai2018digital,deepwatermarking2019aaai} have been proposed. In \cite{uchida2017embedding,nagai2018digital}, a special weight regularizer is leveraged so that the distribution of model weights can be resilient to attacks such as fine-tuning and pruning. However, they only work in a white-box way and need to know the original network structure and parameters for retraining. 
By contrast, Adi \textit{et al.}\cite{adi2018turning} establish a tracking mechanism for watermarking DNN in a black-box way, which uses a particular set of inputs as the indicators and lets the model deliberately output specific incorrect labels. Despite their success, most of these methods only concentrate on simple modification-based attacks like fine-tuning. In \cite{szyller2019dawn}, Szyller \textit{et al.} first consider the more challenging surrogate model attack but still only focus on the classification task. Recently, the work of \cite{deepwatermarking2019aaai} starts to consider the watermarking problem for image processing networks and innovatively leveraged spatial invisible watermarking algorithms for model watermarking. 
However, as pointed out in their extension work \cite{deepwatermarking2021pami}, it will totally fail when the attacker utilizes some data augmentation during the surrogate model's training, as the underlying working principle relies on the whole-image watermark consistency. Our method is motivated by this method, but designs the new structure consistency to obtain augmentation robustness.

\vspace{0.4em}
\noindent\textbf{Data Augmentation.}\label{sec2.3}
Data augmentation plays a crucial role in learning better generalized deep neural networks.
Basic data augmentation techniques include rotation, flipping, cropping, and adding noises. 
Depending on whether they affect the original image quality, we divide them into two categories: quality-harmless and quality-harmful. 
For deep image processing tasks, since the attacker wants the surrogate to get high-quality output, we mainly consider the common quality-harmless data augmentation techniques such as flipping, rotation, cropping, and resizing. If the attacker uses quality-harmful ones such as adding noises, the quality of the surrogate model itself will be bad. Nevertheless, we still consider 6 quality-harmful augmentations (namely, noise, blur, hue, saturation, contrast and style transfer) as an ablation study to test the robustness.

\vspace{0.4em}
\noindent\textbf{Image-to-Image Translation.}
Image-to-image translation is a typical image processing task of which the input and output are both images. It is widely adopted in many applications such as edge to the image synthesis, deraining, and X-ray Chest image debone. In recent years, Generative Adversarial Network (GAN)\cite{goodfellow2014generative} has brought significant progress to image-to-image translation. Generally, there are three typical settings: paired, unpaired, and semi-paired. Regarding the pair setting, Isola \textit{et al.} \cite{pix2pix2017} propose a conditional translation framework named ``pix2pix'' to learn the mapping from input to output, which is improved by many following works\cite{choi2018stargan,wang2018high,park2019semantic}. For the unpaired setting, Zhu \textit{et al.} \cite{zhu2017unpaired} leverage the cycle consistency and propose a general unpaired translation framework CycleGAN. In \cite{eusebio2018semi}, a semi-supervised learning algorithm is proposed. Similar as \cite{deepwatermarking2019aaai,deepwatermarking2021pami}, because paired data is more difficult and expensive to collect, and many high-quality deep processing models are also trained in a supervised way, we mainly consider the pairwise translation as the example applications.


\section{Pre-analysis and Motivation}

\noindent\textbf{Recap of the ``whole-image consistency".} Given an input domain $A=\{a_1, a_2, ..., a_n\}$ and a target output domain $B =\{b_1, b_2, ..., b_n\}$,  pairwised deep image processing is to learn a good target model $\mathbf{M}$ so that $\mathbf{M}(a_i)$ can approach $b_i$ under some pre-defined distance metric $\mathcal{L}$:
\begin{equation}
    \mathcal{L}(\mathbf{M}(a_i), b_i) \rightarrow 0.
\end{equation}

For surrogate model attack, it means that given a target $\mathbf{M}$, the attacker does not know its detailed network structure and weights but can access $\mathbf{M}$  to get a lot of input-output pairs. Because attacker may use an input set different from that used by $\mathbf{M}$, we denote the generated input-output pairs as $\{a_1', a_2', ..., a_m'\}$ and $\{b_1', b_2', ..., b_m'\}$ respectively. Then the attacker will use such pairs to train a surrogate model $\mathbf{SM}$. The working principle of \cite{deepwatermarking2019aaai,deepwatermarking2021pami} is based on the hypothesis that if $\mathbf{SM}$ can learn a good mapping between $\{a_1', a_2', ..., a_m'\}$ and $\{b_1', b_2', ..., b_m'\}$, then if a unified watermark $\delta$ is added to all the output $b_i'$, $\mathbf{SM}$ will also absorb $\delta$ into its output, which can be extracted out for forensics. This is based on the fact that:
\begin{equation}
\begin{aligned}
    \mathcal{L}(\mathbf{SM}(a_i'),b_i')\rightarrow 0 &\Leftrightarrow \mathcal{L}(\mathbf{SM}'(a_i'), b_i' + \delta)\rightarrow 0 \\
   \mbox{when} \quad \mathbf{SM'} &= \mathbf{SM} + \mathbf{\delta}.
 \end{aligned}
\end{equation}
Because of the fitting and loss minimization property of deep networks, $\mathbf{SM}$ can be easily learned to be $\mathbf{SM'}$ by adding a skip connection $\delta$. Since $\delta$ is essentially a unified watermark image embedded in the network outputs, we call it ``whole-image consistency".

\begin{figure}[]
    \centering
    \includegraphics[width=0.95\linewidth]{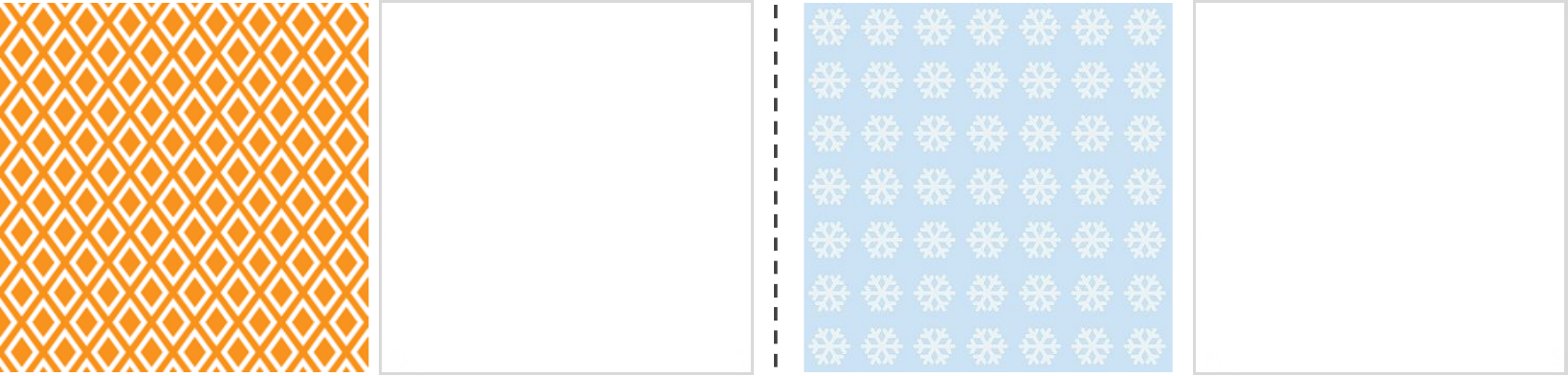}
    \caption{Two repetitive watermark patterns tried to preserve the whole-image consistency in the baseline method. But still no watermark can be extracted from the surrogate model's outputs.}
    \label{fig:repeat}
    \vspace{-2em}
\end{figure}

 \vspace{0.4em}
\noindent\textbf{Fragility to Data Augmentation.} 
Despite the effectiveness,  it has a serious limitation as admitted in \cite{deepwatermarking2021pami}, \ie, the above ``whole-image consistency'' is not robust to data augmentation techniques, which are commonly used in training DNNs. Because data augmentation will destroy the underlying watermark consistency, which is the basis of \cite{deepwatermarking2019aaai,deepwatermarking2021pami}. To preserve the consistency, one intuitive way is to use repetitive watermark patterns as shown in \Fref{fig:repeat} instead of one simple logo image, and train the framework with different augmentation operators. However, we find it still does not work and no watermark pattern can be extracted from the surrogate model's outputs. Because even though the watermark pattern is repetitive, it will still change during the augmentation process, \eg position shift during cropping and orientation change during rotation.

In fact, we will show that the whole-image consistency is indeed methodologically difficult to hold under data augmentation. Specifically, denote the data augmentation operation of each $\{a_i', b_i'+\delta\}$ as $T_i$, the surrogate model $\mathbf{SM}$ trained with augmentation is to learn the mapping between the domain $\{T_1(a_1'), T_2(a_2'), ..., T_m(a_m')\}$ and $\{T_1(b_1'+\delta), T_2(b_2'+\delta), ..., T_m(b_m'+\delta)\}$. Let us simplify the explanation by assuming $T_i$ to be a linear operation, \ie, $T_m(b_m'+\delta) = T_m(b_m') + T_m(\delta)$. For pair-wised image processing, since there exists underlying content relationship between $a_i'$ and $b_i'$, if the same constant $T_0$ ($T_i=T_0$) is applied to all the $(a_i', b_i')$, once the target model $\mathbf{M}$ can learn such a mapping relationship, it should be feasible for $\mathbf{SM}$. However, \textbf{\emph{if different $T_i$ is used for different $i$ and $\delta$ is not content-related to $a_i'$ or $b_i'$, then $T_i(\delta)$ will lose its consistency across different $i$ and is not related to $a_i', b_i'$ either}}. In this case, it is almost impossible for $\mathbf{SM}$ to learn $\delta$ into its output anymore. This is because, without the consistency constraint or content relationship, given $T_i(a_i)$, there is no information available for $\mathbf{SM}$ to predict what $T_i(\delta)$ looks like, thus $\mathbf{SM}$ directly regards it as independent noise and ignores it by considering the whole training set.   

 \vspace{0.4em}
\noindent\textbf{Structure Consistency.} As analyzed above, if we want $\mathbf{SM}$ to absorb the watermark $\delta$, $\delta$ must be able to keep its consistency under data augmentation. A trivial solution is to let $\delta$ be a pure-color image with constant pixel values. However, such a pure-color watermark image is unfriendly to the convolutional watermark extracting network. Because if the extracting network needs to extract such a constant $\delta$ out for different watermarked images, the convolutional weights will be learned to all zeros while only bias term being non-zeros. In this way, even given an unwatermarked image, it will also output $\delta$ too, which loses the forensics meaning. Therefore, we resort to a more advanced way: making the watermark pattern consistent with image structures.

It is inspired by the observation that some global structures like edges or some local semantic structures like ``eyes" of the face are content-related and can keep their physical meaning under the common data augmentation techniques, we call this type of consistency  ``structure consistency". By further encoding the watermark information into specific color values and filling them into these consistent structures, we can generate structure-aligned watermark $\delta_i$ for each $a_i', b_i'$. During the augmentation $T_i$, $\delta_i$ will adaptively change along with $a_i', b_i'$ and keep its alignment with  structures of $a_i', b_i'$. Therefore, it is still possible for $\mathbf{SM}$ to absorb $\delta_i$ based on such structure consistency. 


\begin{figure*}[t]
\centering
\includegraphics[width=0.95\linewidth]{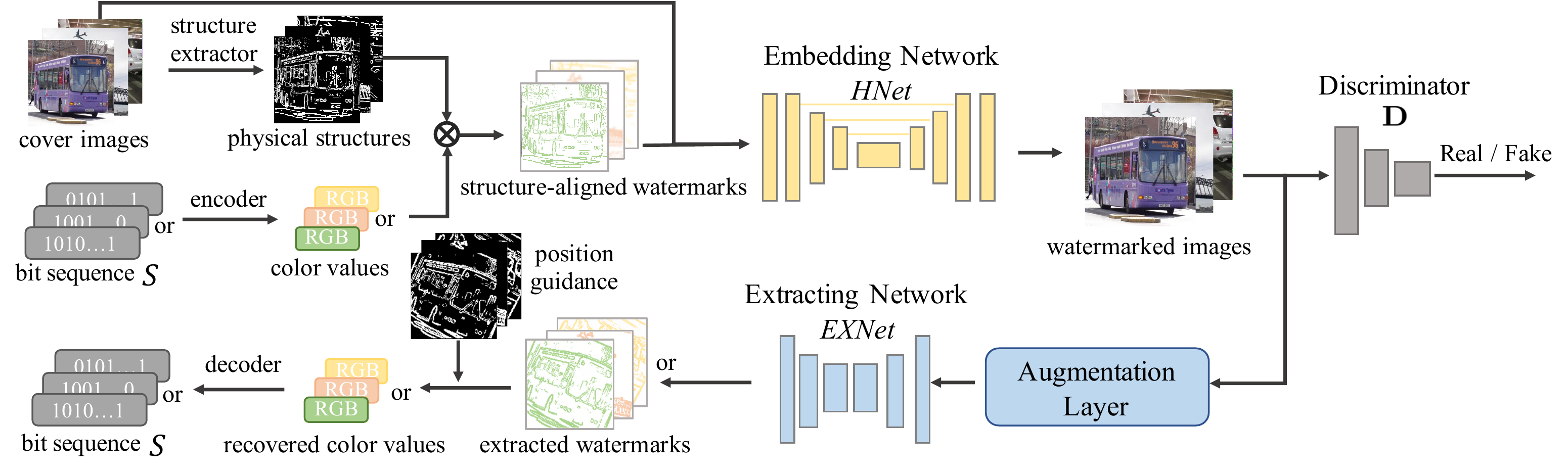}
\caption{The proposed structure-aligned model watermarking framework, which consists of four modules: watermark bit encoder and decoder, one embedding network $\mathbf{\mathit{HNet}}$ and an extracting network $\mathbf{\mathit{EXNet}}$, where the augmentation layer is inserted between $\mathbf{\mathit{HNet}}$ and $\mathbf{\mathit{EXNet}}$. For better performance, a discriminator network $\mathbf{D}$ is also appended. }
\label{fig:system_pipeline}
\end{figure*}

\section{Structure-aligned Model Watermarking}
\noindent\textbf{Overview.} Based on the above structure consistency analysis, we propose the structure-aligned \zj{model} watermarking algorithm in \Fref{fig:system_pipeline}. The basic goal is to learn a good embedding network $\mathbf{\mathit{HNet}}$ and a corresponding extracting network $\mathbf{\mathit{EXNet}}$. $\mathbf{\mathit{HNet}}$ is responsible to embed the structure-aligned watermarks into the cover images to generate watermarked images while $\mathbf{\mathit{EXNet}}$ is responsible to extract the embedded watermarks out. To make $\mathbf{\mathit{HNet}}$ and $\mathbf{\mathit{EXNet}}$ robust to different augmentation operations, an augmentation layer is inserted between them and jointly trained. After the training, given a target model to protect, we feed its output to $\mathbf{\mathit{HNet}}$ before exposing to the public. In such a way, the outputs obtained by the attackers are watermarked. And if one surrogate model is trained with such input and watermarked output pairs, $\mathbf{\mathit{EXNet}}$ can still  extract the target watermarks out from surrogate model's outputs for forensics. Besides, one bit encoder and decoder are leveraged to encode/decode the watermark bits respectively. Below we will elaborate each part in details. For ease of presentation, we will use $b_i$ below as the substitute of $b_i'$. 

 \vspace{0.4em}
\noindent\textbf{Watermark Bit Encoder and Structure Extractor.} \label{region_code} We propose to fill constant RGB values into these structures to ensure consistency in the physical structures during the augmentation process. Taking the common 8-bit color space as an example, the value range of each color channel (``Red", ``Blue,"  and ``Green") would be $[0, 255]$. Assuming the color step used for encoding is $t$, then the total number of possible pixel values $n$ equals  $\frac{255}{t}*\frac{255}{t}*\frac{255}{t}$ (255 left as unwatermarked indicator). Therefore, the max watermark bit sequence length is $log_2(n)$. In real applications, the watermark bit sequence $\mathcal{S}$ represents the IP information we want to embed, such as company name, model ID and version.

Given the $\mathcal{S}$, we can use some simple mathematical encoding schemes \zj{(eg., hash functions)} to map $\mathcal{S}$ into one specific color value $\mathcal{C}_i$. And the detailed physical structure format may be different depending on the specific task. For example, we can use the global edges for general natural images and local semantic regions like ``eyes" or ``noses" for face images. In the following experiments, we will try three different types of physical structures to demonstrate the generality of our method. By default, we use the well-known Sobel edge algorithm to extract the global edges as the physical structure.

 \vspace{0.4em}
\noindent\textbf{Structure-aligned Model Watermarking.} After getting the encoded color $\mathcal{C}_i$ and structure map $\mathcal{M}_i$, we fill $\mathcal{M}_i$ with $\mathcal{C}_i$ to produce a structure-aligned watermark $\mathcal{W}_i$:
\begin{equation}
    \mathcal{W}_i = \mathcal{C}_i \otimes \mathcal{M}_i.
\end{equation}
Here, $\otimes$ means filling $\mathcal{C}_i$ into the regions of $\mathcal{M}_i$ whose mask values are 1 and filling a blank color (R:255,G:255,B:255) otherwise. As we want $\mathbf{\mathit{HNet}}$ be capable of handling different $\mathcal{C}_i$ rather than use an independent $\mathbf{\mathit{HNet}}$ for each possible $\mathcal{C}_i$, we randomly sample different $C_i$ during training. 

After obtaining $\mathcal{W}_i$, we concatenate it with the original cover image $b_i$ along the channel dimension and feed them into $\mathbf{\mathit{HNet}}$ to get the watermarked image $b_i^w$. To ensure the robustness to different augmentation operators $\{T_1, ..., T_k\}$, $b_i^w$ will be randomly processed by one or multiple augmentation operators before being fed into the extracting network $\mathbf{\mathit{EXNet}}$. Then, we will recover the hidden color values from the extracted watermark $\mathcal{W}_i'$ by using the physical structure of the watermarked image as the position guidance. Finally, the original watermark bit sequence will be decoded from the recovered color values. 

 \vspace{0.4em}
\noindent\textbf{Network Structures.}
For fair comparison, we follow \cite{deepwatermarking2019aaai} and adopt the UNet \cite{ronneberger2015u} as our $\mathbf{\mathit{HNet}}$. It is an auto-encoder like network structure and adds multiple skip connections between the encoder and decoder part, which is a widely used design in many image translation tasks. For the extracting network $\mathbf{\mathit{EXNet}}$, we also adopt an auto-encoder like network structure. Specifically, three convolutional layers are used as the encoder and one deconvolutional layer along with two convolutional layers are regarded as the decoder. Several residual blocks are further inserted between the encoder and decoder to enhance its learning capacity. To help achieve better visual quality, we leverage one patch discriminator network $\mathbf{D}$ \cite{pix2pix2017} for adversarial training. 

 \vspace{0.4em}
\noindent\textbf{Loss Functions.} 
The training loss consists of two parts: the embedding loss $\mathcal{L}_{H}$ and the extracting loss $\mathcal{L}_{EX}$:
\begin{equation}
    \mathcal{L} = \mathcal{L}_{H} + \lambda*\mathcal{L}_{EX}.
\end{equation}
where $\lambda$ is the hyper parameter to balance their importance.  $\mathcal{L}_H$ is to ensure  the visual quality of watermarked images while $\mathcal{L}_{EX}$ is to ensure that the hidden watermarks can be successfully extracted out. Therefore, a  too large $\lambda$ will cause inferior visual quality but higher extracting success rate, and too small $\lambda$ will obtain high visual quality watermarked images but the hidden watermark would be too weak to be extracted out. 

The embedding loss $\mathcal{L}_{H}$ has two parts: a simple $L$2 loss $\ell_{2}$ and an adversarial loss $\ell_{adv}$, i.e.,
\begin{equation}
\begin{aligned}
    \mathcal{L}_{H} &= \lambda_1 * \ell_{2} + \lambda_2  * \ell_{adv}.
\end{aligned}
\end{equation}
The $L$2 loss $\ell_{2}$ measures the  pixel-wise difference between the input cover image $b_i$  and  the watermarked output image $b^w_i$. That is to say, we want the watermarked images to be visually similar to the original unwatermarked images so that the attacker even cannot know whether the output of the target model is watermarked or not.
\begin{equation}
\begin{aligned}
    \ell_{2} &= \underset{b_i\in \mathbf{B},b_i^w\in \mathbf{B}^w}{\mathbb{E}} \lVert b_i - b^w_i\rVert^2. \\
\end{aligned}
\end{equation}
Here $\mathbf{B}$ and $\mathbf{B}^w$ represent the unwatermarked and watermarked image set respectively. And the adversarial loss $\ell_{adv}$ will encourage the embedding network $\mathbf{\mathit{HNet}}$ to hide watermarks better so that the discriminator $\mathbf{D}$ cannot distinguish its output from real unwatermarked images $b_i$, 
\begin{equation}
\begin{aligned}
    \ell_{adv} &=  \underset{b_i\in \mathbf{B}}{\mathbb{E}}log(\mathbf{D}(b_i)) + \underset{b^w_i\in \mathbf{B}^w}{\mathbb{E}} log(1 - \mathbf{D}(b^w_i)).\\
\end{aligned}
\end{equation}

For effective forensics, besides the requirement that $\mathbf{\mathit{EXNet}}$ can extract the hidden watermarks out from the watermarked images, we also need $\mathbf{\mathit{EXNet}}$ not to extract any watermark out for unwatermarked images. Therefore, the extracting loss consists of two terms: one for watermarked images $\ell_{wm}$ and one for unwatermarked images:
\begin{equation}
\label{eq:wm}
    \mathcal{L}_{EX} = \lambda_3 *\ell_{wm} +  \lambda_4 * \underset{b_i\in \mathbf{B}}{\mathbb{E}}\lVert \mathbf{\mathit{EXNet}}(b_i) - \mathcal{O} \rVert^2,
\end{equation}
where $\mathcal{O}$ represents the constant image with all pixels values as (R:255,G:255,B:255) for unwatermarked images. To balance the loss contributions from the watermarked and unwatermarked regions, an adaptive weight $\lambda_5$ will be added for watermarked regions. Formally, $\ell_{wm}$ is defined as:
\begin{equation}
\begin{aligned}
\ell_{wm} = &\lambda_5 * \underset{b^w_i\in \mathbf{B}^w}{\mathbb{E}} \lVert \mathbf{\mathit{EXNet}}(b_i^w)  \otimes \mathcal{M}_i - \mathcal{W}_i \rVert^2     \\ 
      & \ + \underset{b^w_i\in \mathbf{B}^w}{\mathbb{E}} \lVert \mathbf{\mathit{EXNet}}(b_i^w) \otimes \overline{\mathcal{M}_i} - \mathcal{O} \rVert^2   .
\end{aligned}
\end{equation}
As defined before, $\mathcal{M}_i$ represents the physical  structure region, $\overline{\mathcal{M}_i}$ is the background region and  $\mathcal{W}_i$ denotes the ground-truth watermark. The weight $\lambda_5$ depends on the ratio of the physical structure area to the total image area. The smaller the ratio, the larger the weight $\lambda_5$. In our implementation, $\lambda_5$ is pre-calculated on a set of training images to ensure $\lambda_5*\sum_i sum(\mathcal{M}_i) \approx \sum_i sum(\overline{\mathcal{M}_i})$.

To enhance the ability of  $\mathbf{\mathit{EXNet}}$ in extracting watermarks from the surrogate models' output, we also add an adversarial training stage as \cite{deepwatermarking2019aaai}. Specifically, one simple surrogate network is used to mimic the attacker's behavior, then $\mathbf{\mathit{EXNet}}$ is finetuned by adding outputs of this surrogate model into its training set. This stage can be regarded as one special augmentation operation from model processing. 

\vspace{0.4em}
\noindent\textbf{Incremental Training Strategy.} 
Unlike \cite{deepwatermarking2019aaai} where the watermarked images are assumed unchanged, watermarked images in our case will be processed under different types of data augmentation. And some augmentation operations will significantly change the original statistics of hidden watermarks and make it more difficult to be extracted. To resist different augmentation operations, we add these augmentation operators into the training process, forming an augmentation layer. 

We find training such a system with all operators together from scratch is challenging. To reduce the learning difficulty, we propose an incremental training strategy by adding augmentation operators one by one into training until the previous one converges. For the objective loss function, we only use the $\ell_2$ loss term in $\mathcal{L}_H$ to constrain the $\mathbf{\mathit{HNet}}$ until all the augmentation operators are added, and then add the adversarial loss $\ell_{adv}$ to fine-tune the $\mathbf{\mathit{HNet}}$ for achieving better visual quality. In the ablation study, effectiveness of this training strategy will be studied. 

\vspace{0.4em}
\noindent\textbf{Relationship to \cite{deepwatermarking2019aaai}.} Our method and \cite{deepwatermarking2019aaai} both focus on the model IP protection problem for deep image processing networks and surrogate model attack. We also follow the common watermarking framework \cite{zhu2018hidden,baluja2019hiding}: one embedding sub-network for watermark embedding and one extracting sub-network for watermark extraction. There are two significant differences between our method and \cite{deepwatermarking2019aaai}. First, the underlying watermarking methodology is totally different, \ie, \cite{deepwatermarking2019aaai} relies on the whole-image consistency, which embeds a unified watermark image into the network outputs and fails if common data augmentations are used. The aforementioned analysis has shown that changing the watermark patterns cannot fundamentally solve this problem. By contrast, our ``structure consistency'' innovatively proposes to embed watermark information into semantic structures, whose consistency are inherently robust to data augmentations. The following experiments will demonstrate the superior robustness of our method. Second, training the watermarking framework with diverse augmentation operators is not a trivial task, and we design a new incremental training strategy to achieve the convergence. We tried to integrate the augmentation layer into the original framework of \cite{deepwatermarking2019aaai}, but it does not work. This is because the whole-image consistency will be destroyed during the augmentation process.

\section{Experiments}
The proposed method can be broadly used in many different commercial systems for IP protection, such as medical image processing and remote sensing image enhancement. And the extracted watermarks can be viewed as legal evidence for IP forensics.  Due  to the lack of large public datasets, we tried the two  example image processing tasks (deraining, X-ray Chest image debone) used in \cite{deepwatermarking2019aaai} and a new artistic portrait generation task to demonstrate our effectiveness. Because of the resource and space consideration, we mainly use the derain task for comparison and ablation, more results can be found in supplementary materials. For comparison, as the baseline method \cite{deepwatermarking2019aaai} is the only effective method to date and other traditional watermark types have already been proved ineffective in \cite{deepwatermarking2019aaai}, we only compare our method with \cite{deepwatermarking2019aaai}. Source code will be released upon acceptance.


\vspace{0.4em}
\noindent\textbf{Dataset.} For image deraining, we use 11000 clean images from the PASCAL VOC dataset \cite{everingham2010pascal} and 5000 clean images from the COCO dataset \cite{lin2014microsoft} as the target domain, and use the synthesis algorithm in \cite{zhang2018density} to generate rainy images as the input domain. VOC images are split into three parts: 5000 for the initial training stage, 5000 for the adversarial training stage, and 1000 for testing, while COCO images are used for surrogate model training. For debone, we adopt the rib suppression algorithm proposed by \cite{yang2017cascade} to generate the training pair based on 6500 X-ray Chest images from the chestxray8 dataset \cite{wang2017chestx}. Similarly, for artistic portrait generation, we use APDrawingGAN\cite{yi2019apdrawinggan} for synthesis from 7500 high-quality celebA images\cite{liu2018large}. They are split in a similar way as the deraining task.

\vspace{0.4em}
\noindent\textbf{Hyper-parameters and Augmentation Setting.}
Before adding adversarial loss into the training phase, the default value of $\lambda,\lambda_1,\lambda_3,\lambda_4,$ all equal to 1 and $\lambda_2=0$, and the learning rate is 0.0002. After that, we change the $\lambda_2$ and $\lambda$ to 0.01 and 10 and decrease the learning rate of $\mathbf{\mathit{EXNet}}$ to 0.00002. The color step $t$ is set as 20. Considering the attacker will only use some quality-harmless data augmentation operations to ensure the surrogate model quality, four most popular augmentations are used in the main comparison to simulate the attacker's behavior: flipping, rotation, cropping and resizing. By default, the rotation range is $[-90^\circ, 90^\circ]$, crop size is chosen from $[64,256]$ and resize range is $[1/2,2]$. When training the surrogate model, we choose the augmentation configuration: random rotation from $-30^\circ$ to $30^\circ$, random cropping with size 224 and randomly resizing to 128. 

\vspace{0.4em}
\noindent\textbf{Recovering Color Values.}
Given an extracted watermark, we directly use a  straightforward algorithm to recover the hidden color value: extracting the physical structure of watermarked images as position guidance and calculating the average value in each color channel as the color value.

\vspace{0.4em}
\noindent\textbf{Evaluation Metric.}  PSNR and SSIM are used as the default visual quality metric. For extracting performance, we define the biggest recovered color value error of different color channels as the actual error value and set 10 as the absolute error value threshold \zj{(TH)}. When the error value falls in the range of the threshold, we define it as a successful extraction. And the successful extracting rate (SR) is the ratio of images with successful extraction. Due to the watermarking mechanism difference, we still use the NC value introduced in \cite{deepwatermarking2019aaai} to measure the baseline method. Compared to the NC value, our metric is more strict. 

\begin{figure*}[t]
    \centering
    \includegraphics[width=0.98\linewidth]{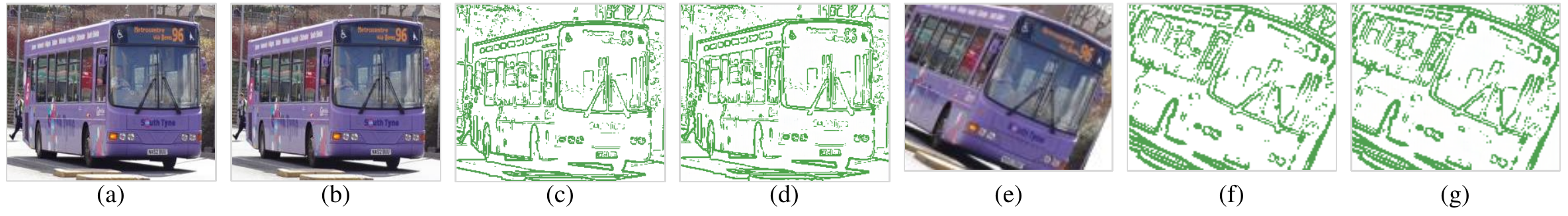}
    \caption{Some visual results of our method: (a) clean image $b_i$, (b) watermarked image $b_i^w$, (c) ground truth watermark $\mathcal{W}_i$, (d) recovered watermark $\mathcal{W}_i'$, (e) $T_i(b_i^w)$ augmented from $b_i^w$, (f) $T_i(\mathcal{W}_i)$ augmented from $\mathcal{W}_i$, (g) recovered watermark from $T_i(b_i^w)$. } 
    \label{fig:visual_results}
\end{figure*}

\begin{table*}[t]
 \footnotesize
 \centering
 \setlength{\tabcolsep}{2.4mm}{
\begin{tabular}{c|c|c|c|c|c|c|c|c|c|c|c}
\hline

\multirow{2}{*}{Pre-processing} & \multirow{2}{*}{Method} & \multicolumn{4}{c|}{Different Network Structures} & \multicolumn{6}{c}{Different Loss Functions}  \\ \cline{3-12} 
                                  &                         & CNet       & Res9       & Res16       & UNet      &  $L$1 & $L$1 + $L_{adv}$ & $L$2 & $L$2 + $L_{adv}$ & $L_{perc}$ & $L_{perc}$+$L_{adv}$ \\ \hline
\multirow{4}{*}{W/O DA} &   \cite{deepwatermarking2019aaai}  &   100\% &100\% & 100\% &100\%          &   100\% &100\% & 100\% &100\%  &86\% &100\% \\  \cline{2-12} 
                                  &    $\textbf{Ours}$     &  100\% &100\% & 100\% &100\%             &   100\% &100\% & 100\% &100\%  &59\% &100\% \\  \cline{2-12} 
               & \cite{deepwatermarking2019aaai} $\dagger$ &   0\%  & 0\% & 0\%  & 0\%                &  0\% & 0\% & 0\% &100\%  &24\% & 0\%  \\ \cline{2-12} 
                                  &  $\textbf{Ours}$  $\dagger$   &  84\%  & 82\% & 84\%  & 45\%      &  54\% & 99\%  & 45\%  & 99\%  & 0\%  & 98\%  \\ \hline
\multirow{4}{*}{With DA}   &  \cite{deepwatermarking2019aaai} &   0\%  & 0\% & 0\%  & 0\%             &   0\% &0\% & 0\% &0\%  &0\% &0\%     \\ \cline{2-12} 
                                  &  $\textbf{Ours}$    &      98\%  & 97\% & 95\%  & 99\%            & 99\% & 97\%  & 99\%  & 96\%  & 57\%  & 97\%  \\ \cline{2-12} 
       &  \cite{deepwatermarking2019aaai} $\dagger$   &   0\%  & 0\% & 0\%  & 0\%          &    0\% & 0\% & 0\% & 0\%  &0\% & 0\%   \\ \cline{2-12} 
                       & $\textbf{Ours}$  $\dagger$   &   0\%  & 0\% & 0\%  & 0\%          &    0\% & 2\%  & 0\%  & 1\%  & 0\%  & 1\%         \\ \hline
\end{tabular}}
\vspace{0.4em}
     \caption{The success rate of resisting surrogate model attack for different network structures and different loss functions without / with data augmentation (DA). $\dagger$ denotes without the adversarial training stage \zj{, and the false positive rates of both methods are 0 for all cases.}}
     \label{tab:robustness}
    \vspace{-0.5em}
\end{table*}

\subsection{Comparison Experiments}

\noindent\textbf{Results of watermarked images and extracted images.} To ensure the watermark embedding network $\mathbf{\mathit{HNet}}$ can embed the structure-aligned watermarks into the cover image $b_i$ and guarantee the watermarked image $b_i^w$ is visually similar to the $b_i$, we first evaluate the PSNR and SSIM values between the watermarked images and the original clean images on the test dataset. Results show that our method can obtain visually indistinguishable watermarked images with the PSNR value as 37.86 and the SSIM value as 0.97. One example visual result is presented in \Fref{fig:visual_results}. It can be seen that our method can extract the hidden watermarks out for both unaugmented and augmented images while guaranteeing high visual quality for watermarked images. Need to note that the end users can only see $b_i^w$ but not $b_i$. 




\vspace{0.4em}
\noindent \textbf{Robustness to different types of surrogate model attack.}
For fair comparison with \cite{deepwatermarking2019aaai}, we follow its setting and evaluate the robustness to surrogate model attack by using different surrogate models with respect to network structures and loss functions. Specifically, four different network structures are used: vanilla convolutional networks only consisting of several convolutional layers (``CNet"), an auto-encoder like networks with 9 and 16 residual blocks (``Res9", ``Res16"), and the aforementioned UNet network (``UNet"); For objective loss function, $L_1$, $L_2$, perceptual loss $L_{perc}$, adversarial loss $L_{adv}$ and their combinations are adopted. By default, $\mathbf{SM}$ model with ``UNet" and L2 loss is leveraged in the adversarial training stage, therefore this configuration can be viewed as white-box attack and all other configurations are black-box attacks. 

For the computation resource consideration, 
we follow \cite{deepwatermarking2019aaai} and conduct controlled experiments to demonstrate the robustness to the network structures and loss functions respectively. Specifically, for the comparison regarding different network structure, the $\mathbf{SM}$ model is only trained with $L2$ loss. And for the loss function comparison,  the $\mathbf{SM}$ model adopts the UNet by default. Below, we consider two different training settings: without data augmentation like \cite{deepwatermarking2019aaai} and with data augmentation.

\vspace{0.4em}
\noindent\textbf{\textit{Without data augmentation.}}
In this ideal setting, the attacker does not pre-process the collected input-output pairs. As shown in \Tref{tab:robustness}, both our method and the baseline \cite{deepwatermarking2019aaai} are robust to different surrogate network structures and loss functions with the adversarial training stage. But without the adversarial stage, our method can still obtain pretty good results for most cases while the baseline \cite{deepwatermarking2019aaai} almost fails. In this sense, our structure consistency is more robust than the whole-image consistency.

\begin{figure}[t]
    \centering
    \includegraphics[width=0.9\linewidth]{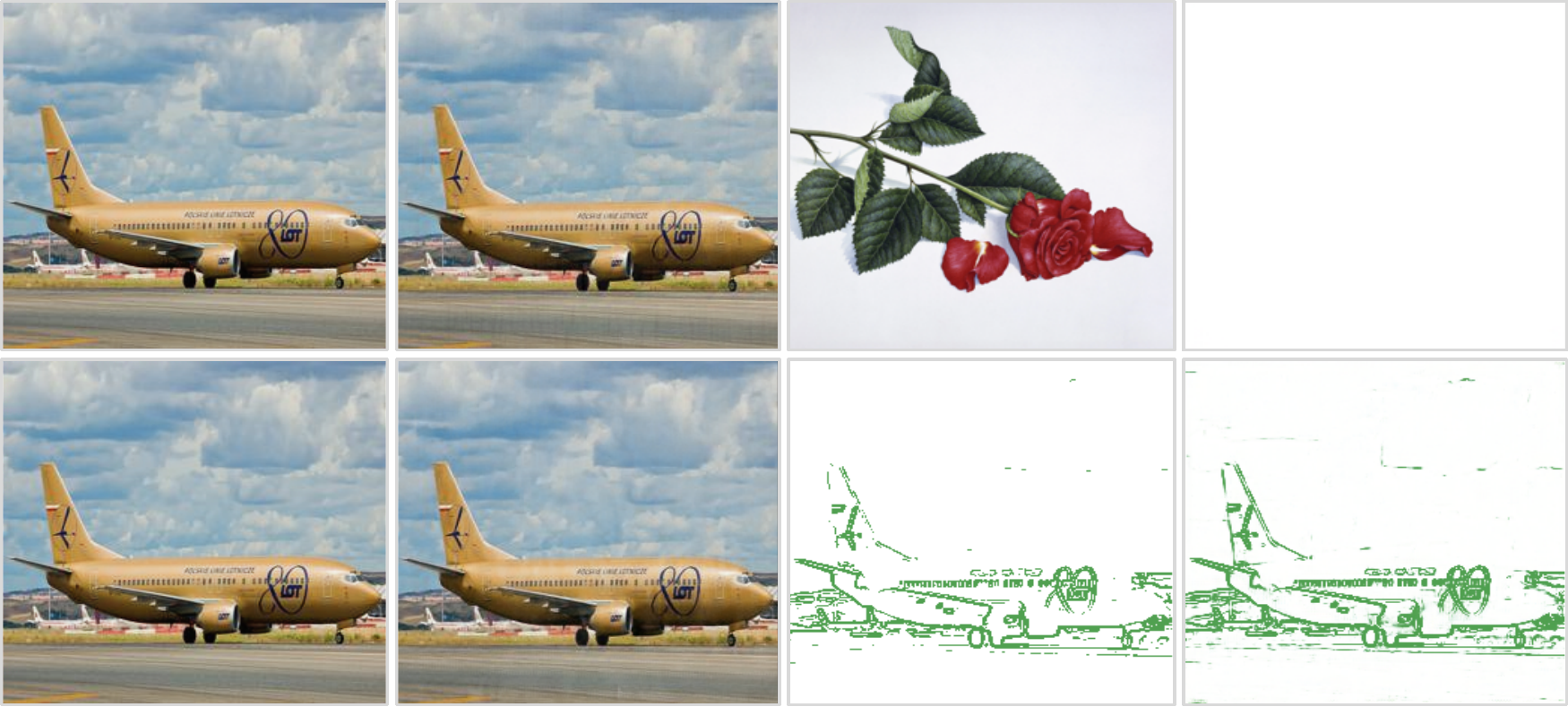}
    \caption{Comparison results of the baseline \cite{deepwatermarking2019aaai} (top) and  our method (bottom). For both methods, from left to right: original watermarked image, surrogate model output, ground truth watermark, and extracted watermark from surrogate model output.}
    \label{fig:vis_bs_cmp}
\end{figure}

\vspace{0.4em}
\noindent\textbf{\textit{With data augmentation.}}
In this more realistic attack scenario, the attacker will utilize data augmentation operators to train the surrogate model $\mathbf{SM}$. As shown in \Tref{tab:robustness},  our method succeeds in most scenarios after adversarial training while the baseline method \cite{deepwatermarking2019aaai} totally fails even after adversarial training, no matter what kinds of network structure or loss function were used. We also observe that the extra adversarial training stage is very important in such challenging data augmentation cases.  In \Fref{fig:vis_bs_cmp}, we provide some visual results about the extracted watermarks from the outputs of the learned surrogate model. Obviously, after data augmentation, the surrogate model of the baseline \cite{deepwatermarking2019aaai} cannot learn the watermark into its outputs anymore.

\begin{figure*}[t]
	\centering
    \includegraphics[width=1\linewidth]{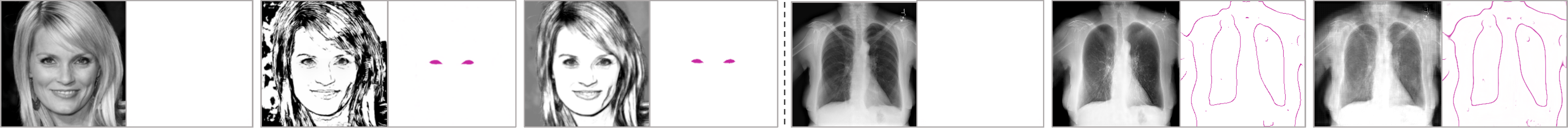}
    \caption{ Example pairs of images and extracted watermarks for the APG and debone tasks. From left to right for both tasks: input domain image $a_i$,  watermarked image $b_i^w$ and surrogate model output. }	
    \label{fig:application}
\end{figure*}

\begin{table*}[t]
\centering
\setlength{\tabcolsep}{2.6mm}{
\footnotesize
\begin{tabular}{c|c|c|c|c|c|c|c|c}
\hline
Settings & Baseline & Noise         & Blur          & Hue           & Saturation    & Contrast      & Style Transfer & Clean    \\ \hline \hline
PSNR     & 32.02    & 31.67 / 31.29 & 31.99 / 31.89 & 32.02 / 32.03 & 31.99 / 31.93 & 32.00 / 31.97 & 31.76 / 31.06 & 32.05 / 32.67 \\ \hline
SR       & 100\%    & 100\% / 100\% & 100\% / 100\% & 100\% / 98\%  & 100\% / 100\% & 100\% / 100\% & 99\% / 99\%    & 100\% / 68\%  \\ \hline
\end{tabular}
}
\caption{The image quality and successful extracting rate of our framework for  surrogate models trained by mixing some augmented data from other augmentation techniques or clean data.  A / B represents the results with 10\% and 50\% mixing ratios, respectively. }
\label{tab:harm_aug}
\vspace{-0.5em}
\end{table*}
 
\subsection{Ablation Study}

\begin{figure}[t]
    \centering
    \includegraphics[width=0.8\linewidth]{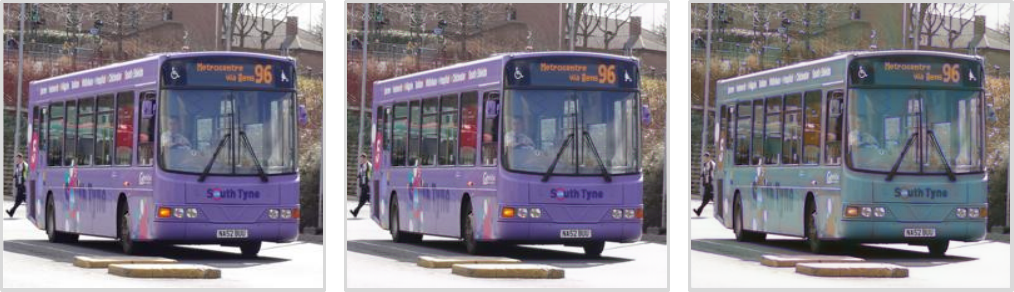}
    \caption{Watermarked images $b_i^w$ comparison with (middle) and without (right) incremental training strategy. And clean images $b_i$ are shown in the 1st column. }
    \label{fig:from_scratch}
\end{figure}
\noindent \textbf{Importance of incremental training strategy.}
As mentioned above, it is very difficult to train the framework with all the augmentation operations and losses from scratch simultaneously. Therefore, an incremental training strategy is adopted. To justify its necessity and superiority, we try to train the framework just from scratch and show the two watermarked images $b_i^w$ in \Fref{fig:from_scratch}. Obviously, this setting suffers from serious color drifting problems. With the proposed incremental training strategy, it works very well.

	

\vspace{0.2em}
\noindent \textbf{Generalization ability.}
Besides the deraining task, we also try another interesting image processing task, called artistic portrait generation (APG). Given a real face image, APG converts it to a pencil drawing style. To demonstrate the generalization ability of structure consistency, we regard the semantic ``eyes" region as the physical structure. Then the extracting network needs to recognize this semantic structure and extract the hidden watermark out automatically, which is more challenging than global edges based physical structures. 
We further apply our framework to the X-ray Chest image debone task, which is also mentioned in \cite{deepwatermarking2019aaai}. We choose another famous  Canny edge algorithm to extract the global edges as the physical structure.  \zj{In \Fref{fig:application}, we provide some visual results of both applications. It can be seen that our framework works very well for different physical structures and is general for different tasks.} More results are given in the supplementary material.

\vspace{0.2em}
\noindent \textbf{Robustness to other augmentation attacks.} As mentioned before, we only consider quality-harmless augmentation by default and assume all the training pairs used by the surrogate model are the output of our target model. But like the arm race, the attacker may train the surrogate model with partial quality-harmful augmented data or self-labeled data to destroy the consistency constraint and remove the watermark. To simulate such behaviors, we mix some watermarked data augmented by \zj{6 representative quality-harmful techniques} and some unwatermarked data into the surrogate model training dataset, respectively. In \Tref{tab:harm_aug}, two mixing ratios (10\%/50\%) are considered. Surprisingly, though the consistency constraint is destroyed in the newly introduced data, our method can still work very well in resisting surrogate model attacks, even when $50\%$ self-labeled clean data is added. 
 Note that we do not retrain our framework here. \zj{More details are given in the supplementary material.}

\noindent \textbf{Robustness to more circumvention attacks.}
Apart from data augmentation attacks, attackers may consider more strategies to remove the model watermark.
First, we consider Neural Cleanse \cite{wang2019neural}, which is famous for reverse-engineering the watermark pattern. But it totally fails because our method is designed in a global and structure-aligned way, which does not fulfill its assumption, \ie, \textit{the  trigger is input-agnostic and static both in location and pattern}).
Second, we assume the attacker collects a small amount of clean (unwatermarked) data pair, and conducts supervised fine-tuning. Results show that even finetuning with a new same-size clean data, our  $\mathbf{\mathit{EXNet}}$ can still work well with 78\% success rate.
Third, we consider the case where the attacker has un-paired clean data and trains the surrogate model with a domain-adversarial loss (watermarked vs. non-watermarked images). In this case, the extracting success rate  degrades to 43\% but it is still acceptable. Moreover, doing this will hurt the surrogate model's performance (PSNR: from 32.3 to 28.9) and make the attack less meaningful. Finally, we consider the robustness of our method to watermark overwriting. Similar to traditional media watermarking, overwriting can be solved by watermark legal agreement. On the other hand, after overwriting, our method can still extract the original watermark out, and the surrogate model performance will degrade a lot.

\section{Conclusion}
Starting from a deep analysis on the model watermarking scheme of \cite{deepwatermarking2019aaai}, we find the fragility of the whole-image consistency is the root cause why this watermarking framework cannot resist the data augmentation attack. To overcome this limitation, we propose a new watermarking methodology,  ``structure consistency", based on which a novel and robust structure-aligned model watermarking algorithm is designed. Experiments demonstrate that the ``structure consistency" can be utilized in both a global and local way, and achieve much better robustness to data augmentation attack and other circumvention attacks.

 \section*{Acknowledgments}
	This work was supported in part by the NSFC Grant U20B2047, 62002334 and 62072421, Exploration Fund Project of University of Science and Technology of China under Grant YD3480002001, and by Fundamental Research Funds for the Central Universities under Grant WK2100000011, and Hong Kong ECS grant 21209119, Hong Kong UGC. Gang Hua is partially supported by National Key R\&D Program of China Grant 2018AAA0101400 and NSFC Grant 61629301. And Jie Zhang is partially supported by the Fundamental Research Funds for the Central Universities WK5290000001. 

{\small
\bibliographystyle{unsrt}
\bibliography{egbib}
}

\end{document}